# Structure and magnetic properties of nanostructured Pd–Fe thin films produced by pulse electrodeposition

G.K. Strukova[1], G.V. Strukov[1], S.I. Bozhko[1], Yu.P. Kabanov[1], I.M. Shmytko[1], A.A. Mazilkin[1], N.A. Sobolev[2], E.R. Zhiteytsev[2,3], A.A. Sukhanov[3], V.K. Voronkova[3], L.R. Tagirov[4]

[1]*Institute of Solid State Physics, Russian Academy of Sciences, 142432 Chernogolovka, Russia*
[2]*Departamento de Física and I3N, Universidade de Aveiro, 3810-193 Aveiro, Portugal*
[3]*Zavoisky Physical Technical Institute, 420029 Kazan, Russia*
[4]*Kazan State University, 420029 Kazan, Russia*

Corresponding author: Dr. G.K. Strukova, Institute of Solid State Physics, Russian Academy of Sciences, 142432 Chernogolovka, Russia; tel.: +7 496 522 2971, FAX: +7 496 524 9701, e-mail: strukova@issp.ac.ru

Nanostructured Pd–Fe thin films with varied Fe content were prepared by electrodeposition technique from organic electrolytes on Cu and brass substrates. The structure and the magnetic properties of the films were investigated prior to post-deposition annealing. The structure of the $Pd_{1-x}Fe_x$ thin film with $x = 0.14$, 0.24, and 0.52 was determined by X-ray diffraction (XRD) and transmission electron microscopy (TEM) as a solid solution of iron in palladium face-centered cubic lattice with the (111) orientation of nanograins relatively to the substrate surface. The films with higher iron concentration, $x = 0.74$, 0.91, have structure of a solid solution based on the body-centered cubic lattice. The average grain size determined by the scanning electron microscopy (SEM) for the first two alloys is 7–10 nm, and for the latter ones it is about 120 nm. The saturation magnetization of the films has linear dependence on the iron content, but coercivity has non-monotonic dependence on $x$, *i.e.* the films with $x = 0.68$ show highest coercivity. The magnetic anisotropy of the samples is studied by ferromagnetic resonance (FMR) spectroscopy.

# 1. Introduction

The $Pd_{1-x}Fe_x$ system is of particular interest due to its fundamental physical properties and possible practical applications because of its high resistance to corrosion. The alloys with $x \approx 0.05$–$0.10$ are weak ferromagnets with $T_{Curie} \approx 100$ K and of interest for superconductor–ferromagnet hybrids studies. In the alloys with $x \approx 0.5$ it is possible to obtain the $L1_0$ phase suitable for perpendicular magnetic recording applications. The alloy with $x \approx 0.7$ shows a shape memory effect induced by an external magnetic field, and is a good choice for sensors. In order to obtain homogeneous magnetic properties at the sub-micron structural level, the nanocrystalline Pd-Fe films are intensively investigated, and new methods for their production are under a search. High vacuum, high temperature and sophisticated equipment are necessary to obtain Pd-Fe films by means of vacuum deposition technique, which makes the product quite expensive. Alternative electrodeposition techniques are more than 10 times cheaper. However, side reactions of the protons electrical reduction and hydrogen emission take place during the electrodeposition process from water-base electrolytes(?). These lead to the hydrogen saturation of the films and, as a consequence, to internal stresses, porosity, surface roughness and inhomogeneity of the film properties [1]. The key idea of our approach is to carry out the reaction in organic aproton-dipolar solvent, which allows to obtain nanocrystalline Pd–Fe films without the drawbacks mentioned above. We further characterize the structure and magnetism of the electrodeposited Pd–Fe films to obtain their physical properties.

# 2. Experimental details

## 2.1. Electrodeposition method

Pd-Fe films with typical thicknesses of 20–100 nm were grown on copper and brass substrates by the pulsed electrodeposition. A sample with an average surface area of about 8 cm$^2$ is used as a cathode, while the anode electrode is an iron or palladium plate. When a sample is inside the electrodeposition cell, a generator passes rectangular current pulses through the electrolyte solution, thereby depositing a metal layer on the sample. The amplitude (50–300 mA) and duration (15–300 ms) of the current pulses are set by a program [2]. The current density was varied from 6.3 to 37 mA/cm$^2$. In this technique, we used a single electrolyte containing palladium and iron ions in an organic aproton-dipolar solvent [3]. This organic solvent forms, together with ammonium chloride, complexion agents for the $Fe^{2+}$ and $Pd^{2+}$ ions, thereby reducing a difference in the redox potentials. This enables one to obtain Pd-Fe alloy coatings in a wide range of compositions. The Fe content in the coating was varied both by modulating the metal concentration in the solution, and the current density as well. For example, the $Pd_{86}Fe_{14}$ film was obtained from a solution containing 43.2 mM/l

Pd, 5.4 mM/l Fe, and 1.0 M/l ammonium chloride with an applied current density of 6.3 mA/cm$^2$. The solution containing 11.7 mM/l Pd, 44.7 mM/l Fe, and 0.9 M/l ammonium chloride with an applied current density of 37.5 mA/cm$^2$ was used to obtain the Pd$_9$Fe$_{91}$ film. The working temperature was kept 50°C in a volume of 80 ml with agitation.

### 2.2. Characterization techniques

The XRD analysis was carried out using Siemens D500 diffractometer. The diffraction patterns were recorded using Cu Kα radiation. The chemical analysis was performed on YXA-5 X-ray microanalyser equipped with LINK AN-10000 analytical system. The microstructure and surface morphology were imaged by SUPRA-50VP SEM. The grain size was defined by TEM using JEM 100CX microscope. The magnetic structure was investigated by the magnetic force microscopy (MFM, Solver ProM). The magnetic properties were measured on a vibrating sample magnetometer (VSM–155, EG&G). The FMR spectra were recorded on Bruker EMX-plus (Kazan) and ESP 300 (Aveiro) spectrometers at the frequency of about 9.5 GHz.

## 3. Results and discussion

X-ray diffraction measurements were performed to identify phase composition of the Fe-Pd thin films. In Fig. 1 the XRD spectra for the films of pure Fe and Pd, and for the Fe-Pd alloys of different composition are presented. The films were deposited on top of copper and brass substrates.

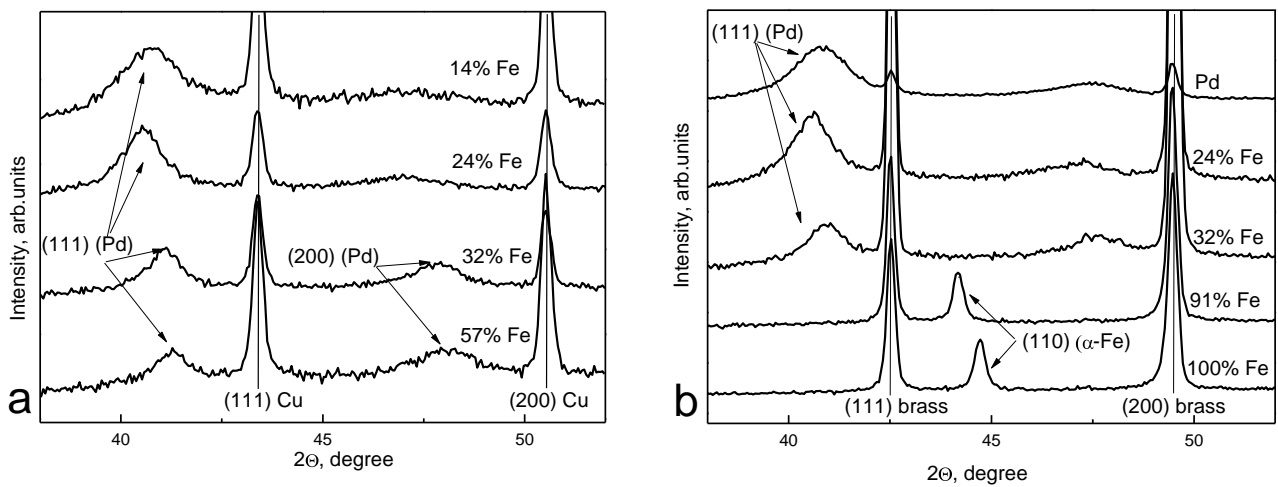

**Fig. 1** X-ray diffraction patterns of Pd-Fe alloys deposited onto copper (a) and brass (b) substrate; the Fe percentage is indicated over the corresponding curve. The percentage content of iron is given as a measure of palladium substitution.

The peaks corresponding to the deposited films are substantially broadened. The analysis of the broadening shows that it takes place due to the small size of the crystallites. The film with 91% Fe is a bcc solid solution of Pd in Fe (here and below the percentage content of iron is given as a measure of palladium substitution). The films with 24, 32 and 57% of Fe consist of a fcc solid solution. The considerable change in the intensity of the (111) and (200) lines of the $Pd_{1-x}Fe_x$ alloy demonstrate the presence of the (111) texture in the grains orientation relatively to the substrate surface. The XRD data show that the Pd- and Fe-related solid solution area for the $Pd_{1-x}Fe_x$ films deposited at 50°C is considerably enlarged as compared to the bulk phase diagram. We consider this result as a size effect of crystallites formed in the films. It is shown that decrease of the grain size can result in a drastic rise of the mutual solubility of the binary system components The XRD results are confirmed by the TEM investigations (see Fig. 2).

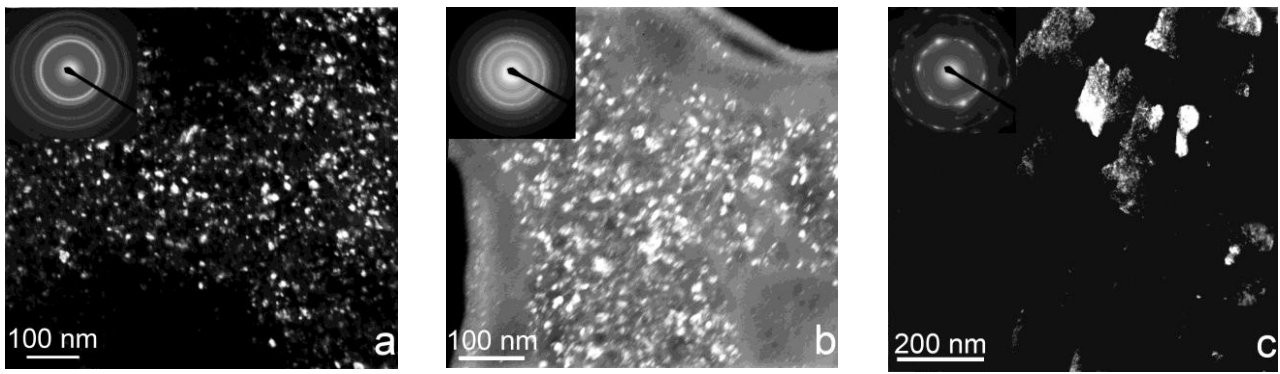

**Fig. 2.** Typical cross-sectional TEM images of the Pd-Fe film with 14 (a), 52 (b) and 74% (c) of Fe. The inserts show the respective electron diffraction pattern.

It can be seen that the alloys with 14 and 52% Fe have the fcc structure, whereas the alloy with 74% Fe is a solid solution with the bcc lattice. The average grain size is about 7–10 nm (14 and 52% Fe), and 120 nm (74% Fe).

In Fig. 3 the SEM images of the Pd-Fe films surface are demonstrated. Grain refinement down to several nanometers is a characteristic feature of the alloys with a low Fe content (Fig. 3a). Globular-shaped clusters are observed on the surface of the films with 52% Fe (Fig. 3b). Their size is about 200–400 nm, and they consist of crystallites with a smaller size. For the films with high iron content (74%) the crystallites of irregular shape with a size of about 200 nm are observed.

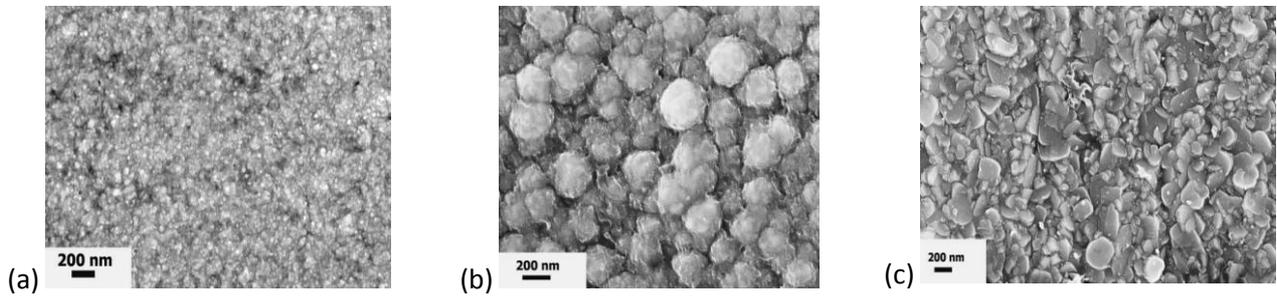

**Fig. 3.** SEM images of films surfaces with different compositions: (a) 14; (b) 52; and (c) 74% of Fe.

A MFM image of this sample with $x = 0.14$ in the as-grown state is shown in Fig. 4. The image clearly shows well-ordered alternating up and down magnetic domain pattern that is caused by the competition between the perpendicular magnetocrystalline anisotropy and the in-plane shape anisotropy. The same domain structure was observed in a $Pd_{50}Fe_{50}$ film grown by molecular beam epitaxy on a single crystalline MgO substrate [4].

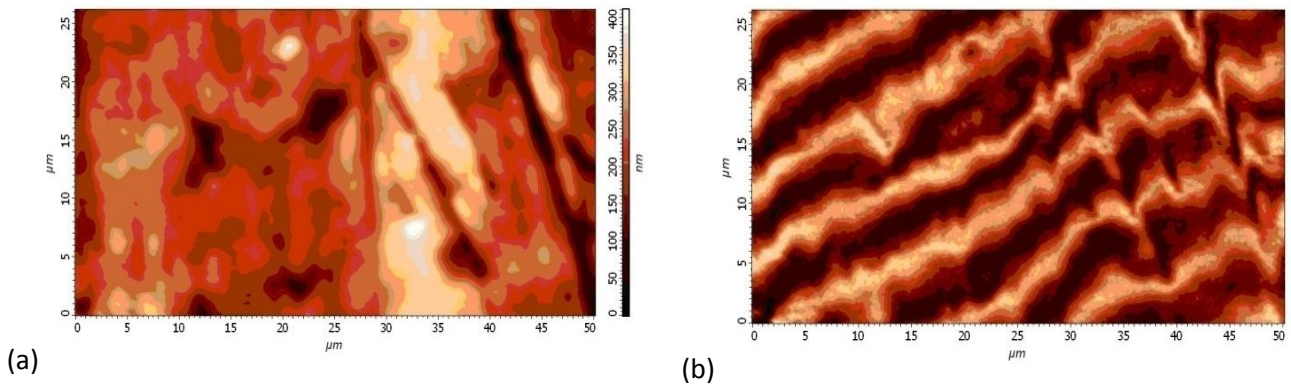

**Fig. 4.** (a) Topography and (b) MFM image of the 100 nm $Pd_{86}Fe_{14}$ film the surface.

The ferromagnetic resonance (FMR) is a very accurate and sensitive technique which allows to determine the magnetic anisotropy fields of thin magnetic films [5]. Experimental FMR spectra for the film containing 14% of iron, with a thickness of 100 nm, are given in Fig. 5; and in Fig. 6 the angular dependence of the in-plane resonance field (Fig. 6a) and the temperature dependence of the 'hard axis' field for resonance (Fig. 6b) are shown. The angular dependences show an anomalously high uniaxial magnetic anisotropy and unconventional triple-line stricture of the FMR spectrum (H1, H2, H3) in the film plane, measured at low temperatures (20K, see Figs. 5a and 6a). As the temperature increases towards room temperature (see Fig. 6b), the magnetic anisotropy dramatically decreases, and two of the three components of the FMR signal (H2 and H3) merge. From the temperature dependences of the resonance field for the low-field component as opposed to the temperature dependence of the two high-field components of the FMR spectrum (see Fig. 6b) it is very probably that a phase with magnetization perpendicular to the film plane is present in the studied sample (see the magnetic hysteresis measurements below). The detailed analysis of the FMR measurements is under intensive development in the fashion of the works [6-8].

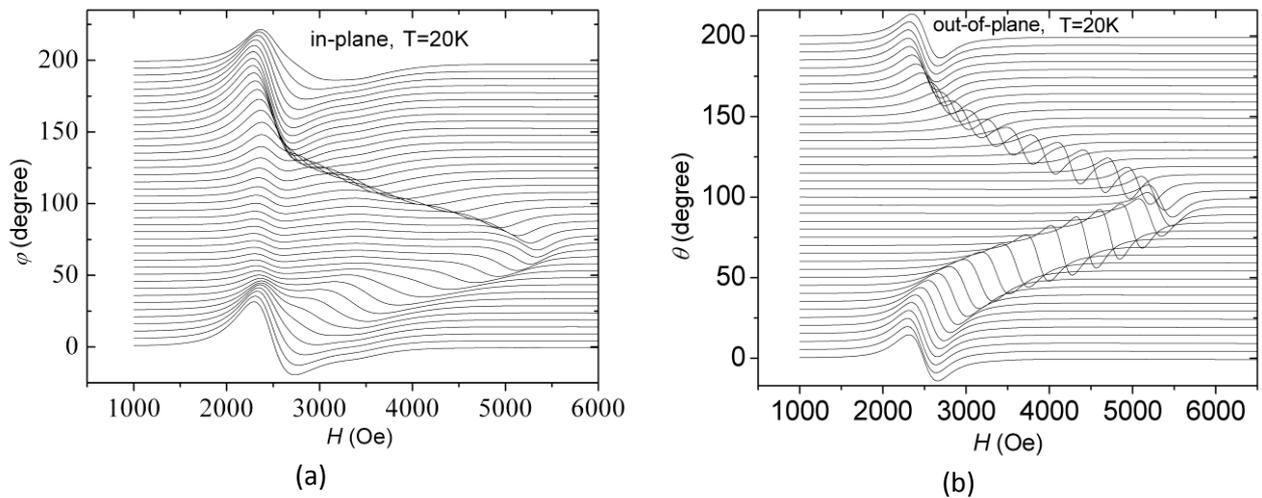

(a)          (b)

**Fig. 5.** Evolution of the FMR spectra upon rotation of magnetic field in the in-plane (a), and out-of-plane ($\varphi = 0^\circ$) (b) geometries for the film with a thickness of 100 nm, and 14% substitution of palladium by iron. The in-plane angle $\varphi$ is counted from direction of the minimal in-plane resonance field ("easy axis"). $\theta = 90^\circ$ corresponds to the external magnetic field orientation perpendicular to the film plane.

Magnetic hysteresis (*M–H*) loops of the deposited Fe-Pd thin films (Fig. 7) were measured to correlate the magnetic properties with the composition and structure (Figs. 1-4).

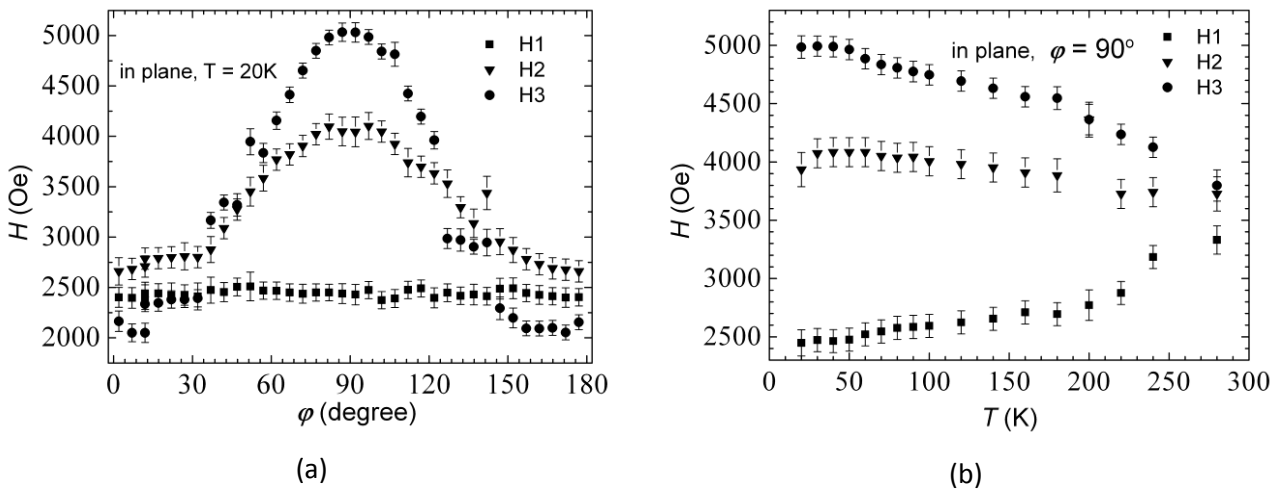

(a)          (b)

**Fig. 6.** (a) In-plane angular and (b) temperature dependences of the resonance field for the $Pd_{86}Fe_{14}$ sample, obtained by decomposition of the FMR spectra on three lines of the Lorentzian shape (H1, H2 and H3).

The saturation magnetization of the films has a linear dependence on the iron content, however, the coercivity does not exhibit any monotonic dependence. The film having 68% of Fe shows the maximum value of the coercivity. The hysteresis loop '4' (for $x = 0.74$) clearly shows presence of the second magnetic phase in the samples.

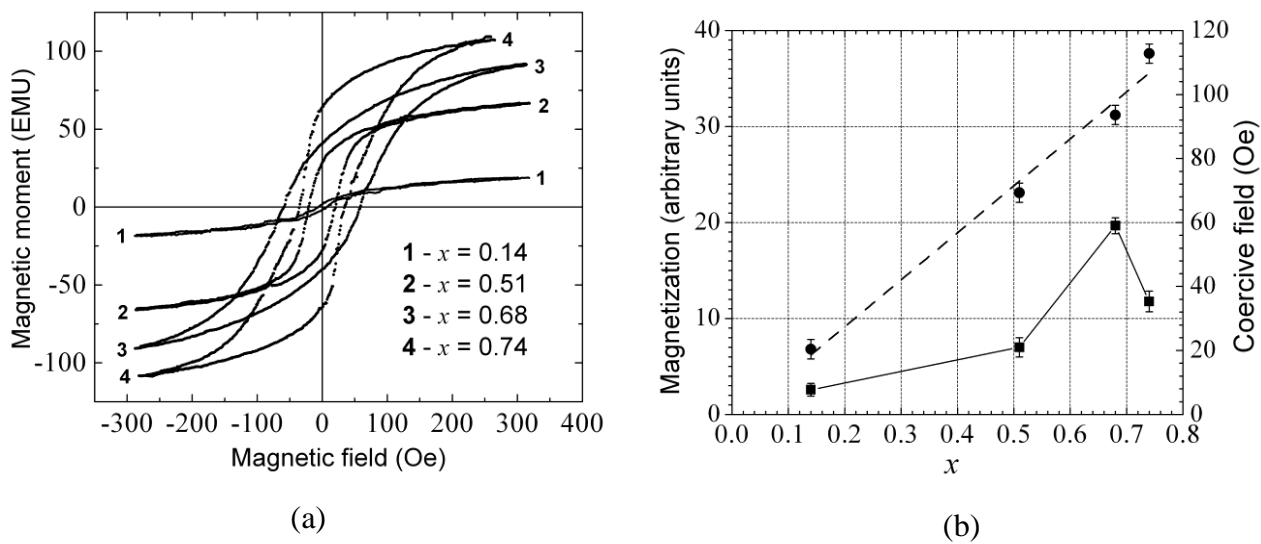

**Fig. 7.** (a) Hysteresis loops for Pd-Fe films with different compositions. (b) Magnetization (solid circles) and coercivity (solid squares) as a function of the iron content $x$ in the Pd-Fe films. Dash line is a liner fit for the magnetization, the solid lines connect experimental point for the coercive field as a function of $x$.

## 4. CONCLUSIONS

In summary, Pd-Fe films of the nanoscale thickness and with a controllable composition were grown by pulsed electrodeposition from organic solutions. The film with 91% of Fe is a solid solution with a bcc lattice textured along the (200) direction. The film with 32 and 24% of Fe is a solid solution with a fcc lattice textured along the (111) direction. According to the XRD, TEM and SEM results, the films containing 50% and more Pd were found to be nanocrystalline with a grain size of 7–10 nm. The films with 52% of Fe consist of 200–400 nm clusters which are composed of finer nanocrystallites. The samples of Pd-Fe films with a Fe content within 12–98% are ferromagnetic at room temperature. The magnetization of the films depends on the Fe content and increases linearly with increasing Fe concentration. MFM and FMR studies revealed in-plane as well as perpendicular anisotropies in the nanocrystalline Pd-Fe films.

### ACKNOWLEDGMENTS

This study was supported by grants within the framework of the Program of Basic Research of the Presidium of the Russian Academy of Sciences "Quantum Physics of Condensed Media."